*Research Article*

# Vulnerabilities Mapping based on OWASP-SANS: A Survey for Static Application Security Testing (SAST)

**Jinfeng Li**

¹Department of Electrical and Electronic Engineering, Imperial College London, London, UK
jinfeng.li@imperial.ac.uk
Correspondence: jinfeng.li@imperial.ac.uk

**Abstract: The delivery of a framework in place for secure application development is of real value for application development teams to integrate security into their development life cycle, especially when a mobile or web application moves past the scanning stage and focuses increasingly on the remediation or mitigation phase based on static application security testing (SAST). For the first time, to the author's knowledge, the industry-standard Open Web Application Security Project (OWASP) top 10 vulnerabilities and CWE/SANS top 25 most dangerous software errors are synced up in a matrix with Checkmarx vulnerability queries, producing an application security framework that helps development teams review and address code vulnerabilities, minimise false positives discovered in static scans and penetration tests, targeting an increased accuracy of the findings. A case study is conducted for vulnerabilities scanning of a proof-of-concept mobile malware detection app. Mapping the OWASP/SANS with Checkmarx vulnerabilities queries, flaws and vulnerabilities are demonstrated to be mitigated with improved efficiency.**

## 1. Introduction

With the prevalence of Internet of Things devices [1] and unprecedented flows of data [2] in the 4G to 5G revolution [3–5] at an exponential pace, the security of web and mobile applications is being increasingly challenged and has gained considerable research interest underpinning a wide variety of industries beyond banking, financial services and insurance (BFSI), such as e-commerce [6], healthcare, telecommunications [7], media, entertainment, retail, education, as well as government and national defense. In this respect, there is arguably an ongoing need for investing massively in the application security sector which enables technological advances for a smarter world. Valued USD 4 billion in 2019, the sharply growing global market in application security is projected to reach USD 9 billion by 2022, and USD 15.25 billion by 2025 at a compound annual growth rate of 25% [8–10].

The amount of easily downloaded mobile applications is constantly on the increase meaning that mobile phones are increasingly vulnerable to malware and other malicious code [10]. Currently, the use of mobile anti-malware systems is not widespread with customers complaining that advertisements and irritating notifications discourage them from using scanners. A customisable anti-malware application is developed in this work which scans APK code files from all other downloaded applications and uses machine learning algorithms to identify potentially malicious code, providing an advert-free experience with only necessary notifications.

Note that embedding security into an application development lifecycle (DLC) encompasses a set of different techniques [11] and assessments at different stages, e.g. Static Application Security Testing (SAST) [12] at an early stage of DLC, and Dynamic Application Security Testing (DAST) [13] at testing and operation stages. SAST scans source code like a white box testing from the inside out, while DAST implements black box testing of the runtime behavior while executing it from the outside in. Comprehensive application security solutions are highly desirable to maximise the coverage of ever-evolving cyberattacks. Among the industry standards of the most critical application security risks, Open Web Application Security Project (OWASP) Top 10 [14] and SANS Common Weakness Enumeration (CWE) top 25 most dangerous software errors [15, 16] are well acknowledged. However, few studies to our knowledge have synced up and mapped the OWASP top 10 with the SANS top 25. This work bridges the gap by performing SAST using Checkmarx, a state-of-the-art source code static scanning tool to identify flaws and vulnerabilities, with the advantages and limitations reviewed in Section 2. A survey on OWASP risk rating methodology is presented in Section 3, followed by the code vulnerabilities mapping into a novel matrix of OWASP Top 10 and SANS top 25 in Section 4 for optimising the checkmark based SAST. A case study incorporating the proposed vulnerabilities mapping is demonstrated for the anti-malware application in Section 5.

## 2. Current Status of SAST based on Checkmarx

In contrast with other application security testing methods (such as DAST which struggles to adequately identify crucial problems within the application layer nor indicate how or where to fix them), SAST based on un-compiled source code analysis offers comprehensive solution into vulnerable patterns and coding flaws from the root up [12]. Specifically, the advantages of Checkmarx-based SAST are summarised below:

- Integrated into delivery pipelines. The SAST service aims at not only providing assurance to security solution consultants, but also enabling developers to write and deliver secure code - this is primarily achieved by integrating the SAST tools into the established development and/or delivery pipeline processes which helps developers discover and fix vulnerabilities long before a project reaches the testing phase.
- Fast and automated. Checkmarx-based SAST technology identifies critical vulnerabilities (e.g. SQL injection and cross-site scripting), allowing instant and relatively accurate feedback on the code with automation, e.g. precisely locating the line number with flaws.
- Low cost. The ability to remediate issues as they arise makes it ideal for integration within the software development lifecycle (SDLC), which saves precious time, remediation efforts and expenses.

However, the current status of SAST tools is susceptible to the following drawbacks [17, 18].

- A large number of false positives and negatives reported, which struggles to confirm that an identified security issue is an actual vulnerability. As a consequence, considerable effort is required for developers to manually identify and remediate the issues.
- A limited percentage of application security flaws can be found automatically. It is still challenging to automatically locate a few types of security vulnerabilities (e.g. authentication problems, access control issues, insecure use of cryptography, etc).
- Limited code coverage, i.e. SAST struggles to locate issues in libraries, configurations, and frameworks, since they are not represented in the code.
- The incapability of reviewing compiled source code and identifying business logic vulnerabilities.

## 3. Methodology in OWASP Risk Rating

Driven by an opensource application security community, the OWASP Top 10 is an industry-standard of the most critical application security risks. The metrics of OWASP is based on a couple of likelihood factors, e.g. weakness prevalence, detectability, exploitability, and technical impact factor [19]. As illustrated in Figure 1 below, the risk rating of the flaws proposed by OWASP is calculated based on two steps. First, average three likelihood factors (i.e. prevalence, detectability, and exploitability), obtaining a likelihood rating. The scale of each risk likelihood factor ranges from 1 (low) to 3 (high). Second, multiply the obtained likelihood rating with a technical impact factor ranging from 1 (low) to 3 (high).

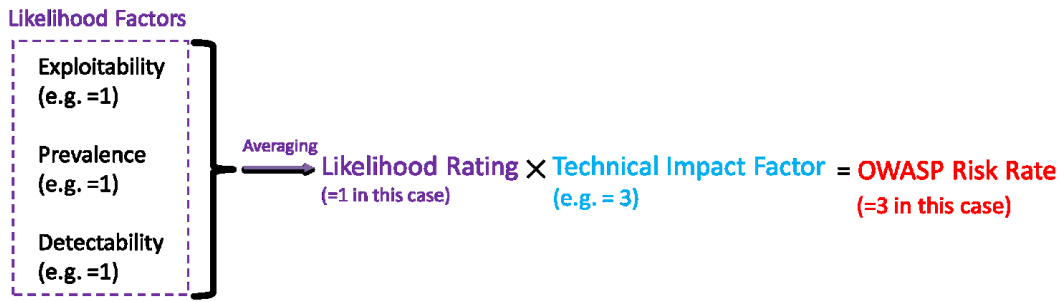

**Figure 1**. Methodology of Calculating the OWASP Top10 Risk Rating.

Based on the above risk rate calculating mechanism, the top 10 vulnerabilities in 2017 [14] are summarised with the corresponding likelihood factors detailed in Figure 2.

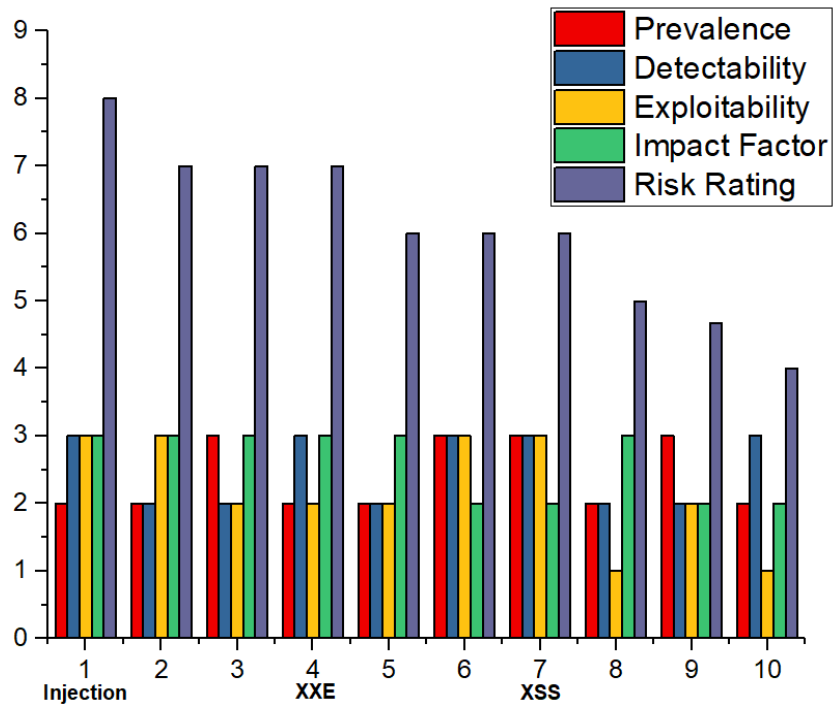

**Figure 2**. Decomposition Analysis of the OWASP Top 10 (Horizontal Axis: 1. Injection Attack, 2. Broken Authentication, 3: Sensitive Data Exposure, 4. XML External Entities (XXE), 5. Broken Access Control, 6. Security Misconfiguration, 7. Cross-Site Scripting (XSS), 8. Insecure Deserialization, 9. Using Components with Known Vulnerabilities, 10. Insufficient Logging and Monitoring).

Over the last few years, attack methods [8–13] have grown with the evolution of fundamental technology and architecture of applications (e.g. JavaScript attacks [20], and attacks on cloud-computing services [21]), and hence the update [14] of OWASP Top 10 as illustrated in Table 1 below.

Table 1. Evolution of OWASP Top 10 from 2013 to 2017 [14].

| | OWASP Top 10 in 2013 | | OWASP Top 10 in 2017 |
|---|---|---|---|
| 1 | Injection | ⇨ | Injection |
| 2 | Broken Authentication and Session Management | ⇨ | Broken Authentication |
| 3 | Cross-Site Scripting (XSS) | ⤵ | Sensitive Data Exposure |
| 4 | Insecure Debit Object References [Merged with 7] | | XML External Entities (XXE) [NEW] |
| 5 | Security Misconfiguration | ⤵ | Broken Access Control [Merged] |
| 6 | Sensitive Data Exposure | | Security Misconfiguration |
| 7 | Missing Function Level Access Control [Merged with 4] | ⤴ | Cross-Site Scripting (XSS) |
| 8 | Cross-Site Request Forgery (CSRF) | ✗ | Insecure Deserialization [NEW] |
| 9 | Using Components with Known Vulnerabilities | ⇨ | Using Components with Known Vulnerabilities |
| 10 | Un-validated Redirects and Forwards | ✗ | Insufficient Logging and Monitoring [NEW] |

## 4. Novel OWASP-SANS Vulnerabilities Mapping

One of the novelty in this work is mapping the co-occurrence of high-profile vulnerability types from both OWASP Top 10 and CWE/SANS Top 25. The obtained matrix is presented in Table 2 according to up-to-date documentation, i.e. 2017 for OWASP [14] and 2019 for CWE/SANS [16].

Table 2. A Novel Vulnerabilities Mapping based on OWASP-SANS/CWE.

| OWASP Rank | OWASP Vulnerability | SANS CWE ID |
|---|---|---|
| 1 | Injection | CWE-78: OS Command Injection (Improper Neutralization of Special Elements used in an OS Command) <br> CWE-89: SQL Injection <br> CWE-94: Code Injection <br> CWE-434: Unrestricted Upload of File with Dangerous Type <br> CWE-494: Download of Code Without Integrity Check <br> CWE-829: Inclusion of Functionality from Untrusted Control Sphere |
| 2 | Broken Authentication | CWE-306: Missing Authentication for Critical Function <br> CWE-307: Improper Restriction of Excessive Authentication Attempts <br> CWE-798: Use of Hard-coded Credentials <br> CWE-807: Reliance on Untrusted Inputs in a Security Decision <br> CWE-862: Missing Authorization <br> CWE-863: Incorrect Authorization |
| 3 | Sensitive Data Exposure | CWE-311: Missing Encryption of Sensitive Data <br> CWE-319: Cleartext Transmission of Sensitive Information |
| 5 | Broken Access Control | CWE-73: External Control of File Name or Path <br> CWE-285: Improper Authorization |
| 6 | Security Misconfiguration | CWE-250: Execution with Unnecessary Privileges <br> CWE-676: Use of Potentially Dangerous Function <br> CWE-732: Incorrect Permission Assignment for Critical Resource |
| 7 | Cross-Site Scripting (XSS) | CWE-79: Improper Neutralization of Input During Web Page Generation (Cross-Site Scripting) |
| 8 | Insecure Deserialization | CWE-134: Use of Externally Controlled Format String |
| 9 | Using Components with Known Vulnerabilities | CWE-190: Integer Overflow or Wraparound <br> CWE-327: Use of a Broken or Risky Cryptographic Algorithm <br> CWE-759: Use of a One-way Hash Without a Salt |

**5. SAST Demonstration on a Proof-of-concept Malware Detection Prototype**

A mobile antivirus software prototype for an Android phone is developed, targeting the functionality of scanning a phone for known flaws and detecting unknown vulnerabilities. The main functional requirements are summarised as follows:

- Customise scan schedule or force immediate scans.
- Monitor potential incoming threats before they are downloaded onto the device.
- Quarantine or block applications that are high-risk and vulnerable.
- View past trends in found vulnerabilities.
- Learn about the dangers of leaving a mobile phone insecure and other cyber-threats.
- Advert-free experience with only necessary notifications.

Checkmarx is employed to perform SAST on the Bitbucket source repository, examining the blueprint of the application without executing the code. By carefully investigating file locations that reported by the Checkmarx vulnerability queries, we observe that 90% of the issues originate from the externally developed libraries used in the python framework (Flask and TensorFlow), which are considered out of scope (marked yellow in Table 3 below) and filtered for a rescan. Only the file internally developed (marked in red below) needs more remediation attention at this stage. The statistics of the initial scanning covering the whole Bitbucket source repository and the rescanning excluding the aforementioned external libraries are reported in Table 4.

**Table 3**. Checkmarx Initial Scan of the Whole Repository including all Libraries.

| High and Medium Vulnerable Files (all libraries scanned) | Number of Identified Issues |
|---|---|
| Android-Malware-Detection-Appmaster/REST_API/venv/lib/python2.7/sitepackages/pip/_vendor/cachecontrol/controller.py | 28 |
| Android-Malware-Detection-Appmaster/ML/preprocess_datasets/create_opcodes_from_apk.py | 13 |
| Android-Malware-Detection-Appmaster/REST_API/venv/lib/python2.7/site.py | 8 |
| Android-Malware-Detection-Appmaster/REST_API/venv/lib/python2.7/sitepackages/pip/_vendor/pkg_resources/__init__.py | 5 |
| Android-Malware-Detection-Appmaster/REST_API/venv/lib/python2.7/sitepackages/pkg_resources/__init__.py | 5 |
| Android-Malware-Detection-Appmaster/REST_API/venv/lib/python2.7/sitepackages/pip/_vendor/distlib/compat.py | 4 |
| Android-Malware-Detection-Appmaster/REST_API/venv/lib/python2.7/sitepackages/pip/_vendor/urllib3/contrib/ntlmpool.py | 4 |
| Android-Malware-Detection-Appmaster/REST_API/venv/lib/python2.7/sitepackages/pip/_vendor/distlib/_backport/sysconfig.py | 3 |
| Android-Malware-Detection-Appmaster/REST_API/venv/lib/python2.7/sitepackages/pip/_vendor/retrying.py | 2 |
| Android-Malware-Detection-Appmaster/REST_API/venv/lib/python2.7/sitepackages/wheel/signatures/djbec.py | 2 |

**Table 4**. Scanning Statistics of the Bitbucket Source Repository.

|  | Lines of Code Scanned | Scan Time | Files Scanned | Coding Language |
|---|---|---|---|---|
| Initial Scan | 137747 | 1h:56m:42s | 605 | Python and JS |
| Rescan | 3116 | 0h:01m:01s | 198 | Python and JS |

Rescan results are analysed in Figures 3 and 4, with 17 vulnerabilities reported in total, exhibiting varying degrees of severity (categorised in high, medium and low). Checkmarx vulnerability queries are mapped with the proposed matrix of OWASP Top 10 and SANS CWE in Table 5, producing a state-of-the-art vulnerabilities matrix guiding application development teams and application security consultants for code remediation.

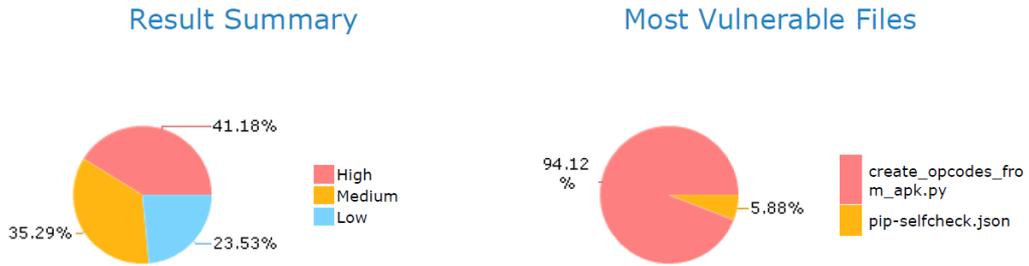

**Figure 3**. Checkmarx Rescanning Results Summary and Locations of the Most Vulnerable Files.

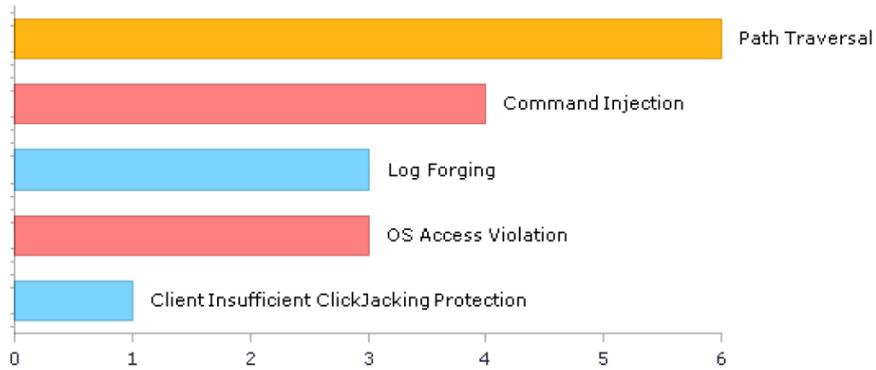

**Figure 4**. Checkmarx Rescanning Results of the Top 5 Vulnerabilities.

**Table 5**. OWASP-SANS Vulnerabilities Mapping with Checkmarx Vulnerability Queries.

| Checkmarx Vulnerability Queries for Antivirus Malware Detection App Demo | OWASP Top 10 (2017) | SANS CWE ID |
|---|---|---|
| OS Access Violation (high) | A5: Broken Access Control | CWE-73 CWE-285 |
| Command Injection (high) | A1: Injection | CWE-78 |
| Code Injection (library scan: out of scope) | | CWE-94 |
| Filtering Sensitive Logs (library scan: out of scope) | A3: Sensitive Data Exposure | CWE-311 CWE-319 |
| Path Traversal (medium) | A5: Broken Access Control | CWE-73 CWE-285 |
| Insecure Randomness (library scan: out of scope) | A6: Security Misconfiguration | CWE-250 CWE-676 CWE-732 |
| Information Exposure Through an Error Message (library scan: out of scope) | | |
| Log Forging (low) | A1: Injection | CWE-494 |
| Client Insufficient Clickjacking Protection (low) | - | - |
| Password In Comment (library scan: out of scope) | A3: Sensitive Data Exposure | CWE-311 CWE-319 |

The above flaws (7 highs and 6 mediums as mentioned in Figure 4) are remediated accordingly incorporating the matrix of Checkmarx queries, OWASP, and SANS, as demonstrated by the final-round of Checkmarx scanning result shown in Figure 5. Only 1 low vulnerability remains, indicating that the application development team can prove closure of the main vulnerabilities.

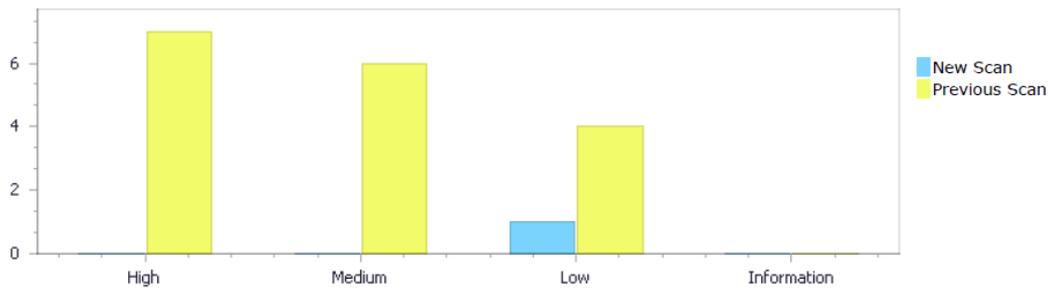

**Figure 5**. Checkmarx Third-round Scanning Results (Compared with the Second-round Scanning).

## 6. Conclusion

This work reviews the recent advances in static application security testing (SAST) and proposes a novel matrix of vulnerabilities mapping based on synchronising the industry-standard OWASP top 10 vulnerabilities, SANS/CWE top 25 most dangerous software errors, and Checkmarx vulnerability queries. With the produced application security framework, enhanced code integrity is demonstrated for a proof-of-concept malware detection application in Android devices through 3 rounds of Checkmarx-based SAST which assists decision making in flaws remediation and vulnerabilities mitigation. The OWASP-SANS matrix-based security framework pioneered in this work can potentially be integrated with other state-of-the-art SAST scanners to expand the security testing scenarios for mobile and web applications.

## References

[1] Ahmad U., Chaudhary J., Ahmad M. and Naz A.A., "Survey on Internet of Things (IoT) for Different Industry Environments", *Annals of Emerging Technologies in Computing (AETiC)*, vol. 3, no. 3, July 2019, pp. 28–43. Available: http://aetic.theiaer.org/archive/v3/v3n3/p4.html

[2] Guo X. Y. and Li J. F., "A Novel Twitter Sentiment Analysis Model with Baseline Correlation for Financial Market Prediction with Improved Efficiency", in Proceedings of the Sixth International Conference on Social Networks Analysis, Management and Security (SNAMS), Granada, Spain, Oct. 2019, pp. 472–477. Available: https://ieeexplore.ieee.org/document/8931720

[3] Li J. F., Xu H. and Chu D.P., "Design of liquid crystal based coplanar waveguide tunable phase shifter with no floating electrodes for 60–90 GHz applications", in Proceedings of the 2016 46th European Microwave Conference (EuMC), London, 2016, pp. 1047–1050. Available: https://ieeexplore.ieee.org/document/7824526

[4] Li J. F. and Chu D.P., "Liquid crystal-based enclosed coplanar waveguide phase shifter for 54–66 GHz applications", *Crystals*, vol. 9, 12, 650, December 2019. Available: https://doi.org/10.3390/cryst9120650

[5] Li J. F., "Structure and Optimisation of Liquid Crystal based Phase Shifter for Millimetre-wave Applications", Doctoral thesis, University of Cambridge, UK, January 2019. Available: https://doi.org/10.17863/CAM.35704

[6] Miraz M. H. and Ali M., "Applications of Blockchain Technology beyond Cryptocurrency", *Annals of Emerging Technologies in Computing (AETiC)*, vol. 2, no. 1, January 2018, pp. 1–6. Available: http://aetic.theiaer.org/archive/v2/v2n1/p1.html

[7] Peter S. Excell, "The British Electronics and Computing Industries: Past, Present and Future", *Annals of Emerging Technologies in Computing (AETiC)*, vol. 2, no. 3, July 2018, pp. 45–52. Available: http://aetic.theiaer.org/archive/v2/v2n3/p5.html

[8] Medeiros I., Neves N. and Correia M., "Detecting and Removing Web Application Vulnerabilities with Static Analysis and Data Mining", *IEEE Transactions on Reliability*, vol. 65, no. 1, pp. 54–69, March 2016. Available: https://ieeexplore.ieee.org/document/7206620

[9]   Shakdher A., Agrawal S. and Yang B., "Security Vulnerabilities in Consumer IoT Applications", 2019 IEEE 5th Intl Conference on Big Data Security on Cloud (BigDataSecurity), IEEE Intl Conference on High Performance and Smart Computing (HPSC) and IEEE Intl Conference on Intelligent Data and Security (IDS), Washington, DC, USA, 2019, pp. 1–6. Available: https://ieeexplore.ieee.org/document/8819463

[10]  Lin Y., Huang C., Wright M. and Kambourakis G., "Mobile Application Security", *Computer*, vol. 47, no. 6, pp. 21–23, June 2014. Available: https://ieeexplore.ieee.org/document/6838873

[11]  Rafique S., Humayun M., Gul Z., Abbas A. and Javed H, "Systematic Review of Web Application Security Vulnerabilities Detection Methods", *Journal of Computer and Communications*, 2015. Available: http://dx.doi.org/10.4236/jcc.2015.39004

[12]  Yang J., Tan L., Peyton J. and Duer K.A., "Towards Better Utilizing Static Application Security Testing", 2019 IEEE/ACM 41st International Conference on Software Engineering: Software Engineering in Practice (ICSE-SEIP), Montreal, QC, Canada, 2019, pp. 51–60. Available: https://ieeexplore.ieee.org/abstract/document/8804441

[13]  Petukhov A. and Kozlov D., "Detecting Security Vulnerabilities in Web Applications Using Dynamic Analysis with Penetration Testing", Proceedings of the Application Security Conference, 2008. Available: https://www.owasp.org/images/3/3e/OWASP-AppSecEU08-Petukhov.pdf

[14]  OWASP, "OWASP Top 10 - 2017 The Ten Most Critical Web Application Security Risks", Open Web Application Security Project, Available: https://owasp.org/www-pdf-archive/OWASP_Top_10-2017_%28en%29.pdf.pdf

[15]  Howard M., "Improving Software Security by Eliminating the CWE Top 25 Vulnerabilities", *IEEE Security & Privacy*, vol. 7, no. 3, pp. 68–71, May-June 2009. Available: https://ieeexplore.ieee.org/document/5054914

[16]  SANS, "CWE/SANS TOP 25 Most Dangerous Software Errors", SANS Institute, 2019, Available: https://cwe.mitre.org/top25/archive/2019/2019_cwe_top25.html

[17]  Wang Y. and Alshboul Y., "Mobile security testing approaches and challenges", 2015 First Conference on Mobile and Secure Services (MOBISECSERV), Gainesville, FL, 2015, pp. 1–5. Available: https://ieeexplore.ieee.org/document/7072880

[18]  Rafique S., Humayun M., Hamid B., Abbas A., Akhtar M. and Iqbal K., "Web application security vulnerabilities detection approaches: A systematic mapping study", 2015 IEEE/ACIS 16th International Conference on Software Engineering, Artificial Intelligence, Networking and Parallel/Distributed Computing (SNPD), Takamatsu, 2015, pp. 1–6. Available: https://ieeexplore.ieee.org/document/7176244

[19]  Ramadlan M.F., "Introduction and implementation OWASP Risk Rating Management", Open Web Application Security Project, 2019. Available: https://owasp.org/www-pdf-archive/Riskratingmanagement-170615172835.pdf

[20]  Ndichu S., Ozawa S., Misu T. and Okada K., " A Machine Learning Approach to Malicious JavaScript Detection using Fixed Length Vector Representation", 2018 International Joint Conference on Neural Networks (IJCNN), Rio de Janeiro, 2018, pp. 1–8. Available: https://ieeexplore.ieee.org/document/8489414

[21]  Duncan A., Creese S. and Goldsmith M., "A Combined Attack-Tree and Kill-Chain Approach to Designing Attack-Detection Strategies for Malicious Insiders in Cloud Computing", 2019 International Conference on Cyber Security and Protection of Digital Services (Cyber Security), Oxford, United Kingdom, 2019, pp. 1–9. Available: https://ieeexplore.ieee.org/document/8885401